\documentclass[11pt]{article} 
\usepackage{graphicx,floatflt,amssymb,epsf,rotate}  
\usepackage{axodraw4j}
\usepackage{pstricks}
\usepackage{color}  
\def\gtap{\raisebox{-.4ex}{\rlap{$\sim$}} \raisebox{.4ex}{$>$}}

\begin{document}  
\begin{flushright}  
\texttt{RECAPP-HRI-2014-021}\\  
\end{flushright}  
 
\vskip 30pt

%opening
\begin{center}  
{\Large{\bf KK-number non-conserving decays: Signal of $n = 2$
excitations of Extra-Dimensional Models at the LHC}}\\
\vspace*{1cm}  
\renewcommand{\thefootnote}{\fnsymbol{footnote}}  
{ {\sf Ujjal Kumar Dey${}^{1,2}$\footnote{email: ujjal.dey1@gmail.com}},  
{\sf Amitava Raychaudhuri${}^{1}$\footnote{email: palitprof@gmail.com}}
} \\  
\vspace{10pt}  
{\small ${}^{1)}$ {\em Department of Physics, University of Calcutta,  
92 Acharya Prafulla Chandra Road, Kolkata 700009, India}}\\ 
  ${}^{2)}$ {\em Harish-Chandra Research Institute, 
Chhatnag Road, Jhunsi, Allahabad  211019, India} \\   
\normalsize

\end{center}

\begin{abstract} 
In the simplest universal extra-dimension  models Kaluza-Klein
(KK) parity distinguishes the states with odd and even KK-number.
We calculate the coupling of a $2n$-level top quark to a top
quark  and the  Higgs scalar (both $n = 0$ states), absent at the
tree level, which is mediated by strong interactions at
one-loop.  We show that the strength of this coupling is
independent of $n$. We observe that the decay due to this coupling,  which
conserves KK-parity, can be a few per cent of the phase space
suppressed decay to two $n$-level states which proceeds through
tree-level couplings. We explore the prospects of verification of
this result  at the Large Hadron Collider through the production
of a second-level KK top-antitop pair both  of which subsequently
decay to a zero mode top quark/antiquark and a Higgs boson. \\

\vskip 5pt \noindent  
\texttt{PACS Nos:~11.10.Kk, 14.65.Jk, 14.80.Rt  } \\  
\texttt{Key Words:~~Universal Extra Dimension, Kaluza-Klein}  
\end{abstract}

\renewcommand{\thesection}{\Roman{section}}  
\setcounter{footnote}{0}  
\renewcommand{\thefootnote}{\arabic{footnote}}

\section{Introduction}

The results of high energy experiments over the last decades,
culminating in the observation of the Higgs scalar \cite{ATLAS,
CMS} at the Large
Hadron Collider (LHC), have continued to strengthen the
confidence on the standard model (SM). Nonetheless, there are issues
such as the evidence for dark matter and the confirmation of
neutrino mass through several oscillation experiments which
compel us to accept that there is interesting physics lying
beyond the realms of the SM. One direction which has received
significant attention is the possibility that there are more
spacelike dimensions than the usual three -- the
extra-dimensional models. There is a wide variety of options
here: the number of extra dimensions, whether the spacetime
metric is dependent on these dimensions or not, and indeed in the
possible ultraviolet completions of such theories. Here we will
restrict ourselves to the simplest of these models, namely, 
Universal Extra Dimensions (UED).

In UED \cite{acd} besides the standard four-dimensional spacetime
there are additional compact spacelike dimensions which are flat
-- constituting the `bulk' -- and all the SM particles have
exposure to these. We will consider models with only one extra
spacelike dimension which we denote by $y$.  The radius of
compactification, $R$, sets a scale for the KK masses. The
coordinate $y$ runs from 0 to $2 \pi R$. Here particles are
represented by five-dimensional fields. Every such field can be Fourier
expanded and expressed  as a tower of four-dimensional
Kaluza-Klein (KK) excitations specified by an integer $n$, the
zero-mode being the corresponding SM particle. To reproduce the
chiral nature of the zero-mode fermions a  $y \leftrightarrow -y$
symmetry is imposed.  Thus the extra dimension is compactified on
the orbifold $S^1/Z_2$. All tree-level couplings when expressed
in terms of the Kaluza-Klein excitations conserve the
KK-number. 

Usually $1/R$ is significantly larger than the SM scale and the
KK states at the $n$-th level have very nearly the same mass,
$n/R$, for all particles.  Thus the mass spectrum is extremely
degenerate. This degeneracy is removed when the five-dimensional
loop contributions \cite{georgi} to the masses of the KK-states
are included.  To evaluate these contributions it is necessary to
introduce a cut-off $\Lambda$ beyond which some more fundamental
theory is expected to be operational. A common practice is to
choose the mass correction to be zero at this cutoff and to
calculate the finite low energy contribution taking this as the
boundary condition \cite{mUED, mUED2}. A symmetry $y \rightarrow
y + \pi R$ is preserved -- referred to as KK-parity -- and is $=
(-1)^n$ for the $n$-th KK-level. These are the ingredients of
minimal UED (or mUED).

The mUED model is completely specified by the cut-off $\Lambda$
and the compactification radius $R$.  It is known that in mUED
electroweak observables receive corrections which are finite  at
one-loop order \cite{db}. This justifies the comparison of the
predictions of this theory with experimental data and obtaining
bounds on $\Lambda$ and $R$. Thus, from the $(g-2)$ of the muon
\cite{nath}, flavour changing neutral current processes
\cite{chk,buras,desh}, $Z \to b\bar{b}$ decay \cite{santa}, the
symmetry breaking $\rho$ parameter \cite{acd,appel-yee}, and
other electroweak precision tests \cite{ewued, precision} it has
been found that $R^{-1}~\gtap~300-600$ GeV.  A relatively modest
$R^{-1}$ encourages the continuing search for signatures of mUED
at the LHC  \cite{collued} and also at other future facilities
\cite{ILC}. Some of the more recent comparisons of UED with the
data,  including Dark Matter constraints,  can be found
in~\cite{recentued}.   In particular, the LHC results
\cite{LHCued} imply $R^{-1} > 600$ GeV from the multijet and
missing $E_T$ data while searches for dilepton resonances yield
$R^{-1} > 715$ GeV. An analysis \cite{LHCued} of the CMS and
ATLAS missing $E_T$ data in the context of a model with two extra
dimensions sets a limit of $R^{-1} > 600$ GeV at 99\% C.L. With
10 fb$^{-1}$ data at the 14 TeV LHC the reach of $R^{-1}$ will be
extended to 1.1 TeV for $\Lambda R$ = 10 \cite{recentued}. Further, Higgs
boson mass and couplings when examined in the context of mUED
suggest $\Lambda R\sim 6$
\cite{cutoff}.

In this work we examine the loop-induced strong interaction
mediated $t^{(2n)} t^{(0)} H^{(0)}$ vertex which respects
KK-parity but does not conserve KK-number\footnote{KK-number
non-conserving decays of the $H^{(2)}$ have been considered
earlier in the context of Kaluza-Klein dark matter models
\cite{kakizaki}.}.  Our notation is
schematic here and will be sharpened later: $t^{(2n)}$ stands for
any of the several top quark excitations of different chirality
at the $2n$-th level. $t^{(0)}$ and $H^{(0)}$ are the SM top
quark and Higgs boson respectively.  We calculate the strength of
this coupling and use it in mUED to compute the decay rate of a
$2n$-level top quark to a zero mode top quark and a Higgs boson.
For the top quark KK excitation this Yukawa coupling-driven
decay mode will dominate over  decays to other zero mode states,
e.g., those with  weak gauge bosons in the final state. We also
compare this rate with the KK-number conserving decay to a pair
of $n$-level states.  Finally, we explore the prospects of
verifying the theory  at the future runs of the LHC through the
detection of a signal using the pair-production of the 
second-level  KK
top quarks and their subsequent direct decays to zero mode
states.

In the following section, after introducing the notations of the
mUED model the calculation of the $t^{(2n)} t^{(0)}
H^{(0)}$ coupling is given. This is followed by an estimation of
the branching ratio of the decay of the $t^{(2n)}$ state through
this coupling. We then use these results to examine the
possibility of detecting a second-level top-quark at the LHC through
its production and decay to zero mode states.  At the end, we
provide a summary and some concluding remarks.

\section{Coupling of the $2n$-level top quark to zero mode states}
\paragraph*{}

The 5-dimensional fields of UED are usually expressed in terms of
a tower of 4-dimensional KK states. For example, the left- and
right-chiral\footnote{The left- and right-chiral projectors are
$(1 - \gamma_5)/2$ and $(1 + \gamma_5)/2$, respectively.} quark
fields of the $i$-th generation will be written as:
\begin{eqnarray} 
\mathcal{Q}_{i}(x,y)&=&\frac{\sqrt{2}}{\sqrt{2\pi  
R}}\bigg[{\pmatrix{u_i\cr d_i}}_{L}(x)+\sqrt{2}\sum^{\infty}_{n=1}\Big[ 
{Q}^{(n)}_{iL}(x)\cos\frac{ny}{R}+ 
{Q}^{(n)}_{iR}(x)\sin\frac{ny}{R}\Big]\bigg], \label{5ddoub} \\ 
\mathcal{U}_{i}(x,y)&=&\frac{\sqrt{2}}{\sqrt{2\pi 
R}}\bigg[u_{iR}(x)+\sqrt{2}\sum^{\infty}_{n=1}\Big[ 
{U}^{(n)}_{iR}(x)\cos\frac{ny}{R}+ 
{U}^{(n)}_{iL}(x)\sin\frac{ny}{R}\Big]\bigg]. 
\label{5dsingl}
\end{eqnarray} 
The expansion for $\mathcal{D}_{i}(x,y)$, containing $d_{iR}$, is
similar to eq. (\ref{5dsingl}). The fields satisfy
$\mathcal{Q}_{i}(x,y) = -\gamma_5
\mathcal{Q}_{i}(x,-y)$ and $\mathcal{U}_{i}(x,y) = +\gamma_5 
\mathcal{U}_{i}(x,-y)$, $\mathcal{D}_{i}(x,y) = +\gamma_5 
\mathcal{D}_{i}(x,-y)$ which ensure that the zero-modes are the
SM quarks with the correct chirality. For the third generation we
use the notation
\begin{eqnarray}
{Q}^{(n)}_{3L} \equiv \pmatrix{t^{(n)} \cr b^{(n)}}_{L} \;\;,&&
{U}^{(n)}_{3R} \equiv t^{(n)}_{R}\;\;,\;\;
{D}^{(n)}_{3R}  \equiv b^{(n)}_{R}\;\;,\;\;
(n = 0,1, \ldots)\;\;,\nonumber \\
{Q}^{(n)}_{3R} \equiv \pmatrix{T^{(n)} \cr B^{(n)}}_{R} \;\;,
&&{U}^{(n)}_{3L} \equiv T^{(n)}_{L}\;\;,\;\;
{D}^{(n)}_{3L} \equiv B^{(n)}_{L}\;\;,\;\;
(n = 1,2, \ldots).
\label{KKquarks}
\end{eqnarray}
Thus, $t^{(0)}_L,b^{(0)}_L$ are the SM third generation
left-handed quarks while  $t^{(0)}_R,b^{(0)}_R$ are similarly
their right-handed counterparts. 

In UED the mass of the $n$-th level KK excitation is $M_n = n/R$
irrespective of the other properties of the field so long as
$1/R$ is much larger than the zero-mode mass, $m_0$, which arises
through the Higgs mechanism\footnote{We use this approximation
for all states. For the top-quark so long as $1/R \sim$ 1 TeV
this is not a bad approximation for our purpose.}. In mUED higher
order corrections to these masses are included. In our
calculation of the  $t^{(2n)} t^{(0)} H^{(0)}$ coupling we use
the lowest order (i.e., UED) masses of the KK states. However,
when we calculate the decay rates in the next section we do
include the mUED corrected masses.

\setcounter{figure}{0}  
\renewcommand{\thefigure}{\arabic{figure}}  
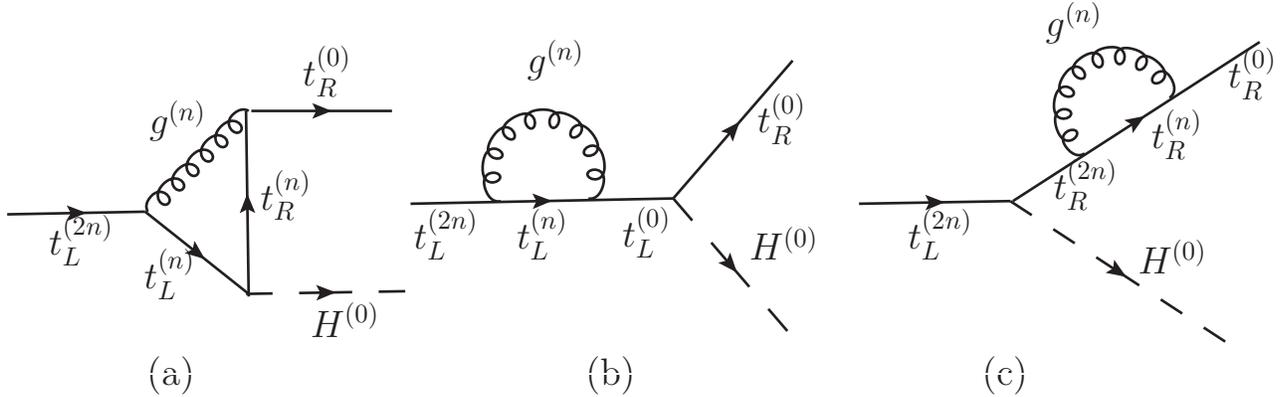
\begin{figure}[tbh]
\begin{center}
\fcolorbox{white}{white}{
  \begin{picture}(498,143) (30,1)
    \SetWidth{1.0}
    \SetColor{Black}
    \Line[arrow,arrowpos=0.5,arrowlength=5,arrowwidth=2,arrowinset=0.2](31,58)(83,59)
    \Gluon(84,59)(122,97){3.25}{6}
    \Line[arrow,arrowpos=0.5,arrowlength=5,arrowwidth=2,arrowinset=0.2](83,59)(122,28)
    \Line[arrow,arrowpos=0.5,arrowlength=5,arrowwidth=2,arrowinset=0.2](122,27)(121,97)
    \Text(84,-10)[lb]{\Large{\Black{(a)}}}
    \Text(250,-10)[lb]{\Large{\Black{(b)}}}
    \Text(400,-10)[lb]{\Large{\Black{(c)}}}
    \Line[arrow,arrowpos=0.5,arrowlength=5,arrowwidth=2,arrowinset=0.2](183,62)(282,64)
    \GluonArc[clock](233.168,74.442)(19.808,-144.714,-391.815){3.25}{8}
    \Line[arrow,arrowpos=0.5,arrowlength=5,arrowwidth=2,arrowinset=0.2](282,64)(327,116)
    \Line[dash,dashsize=10,arrow,arrowpos=0.5,arrowlength=5,arrowwidth=2,arrowinset=0.2](282,64)(325,14)
    \Line[arrow,arrowpos=0.5,arrowlength=5,arrowwidth=2,arrowinset=0.2](352,62)(409,63)
    \Line[dash,dashsize=10,arrow,arrowpos=0.5,arrowlength=5,arrowwidth=2,arrowinset=0.2](409,63)(490,10)
    \Line[arrow,arrowpos=0.5,arrowlength=5,arrowwidth=2,arrowinset=0.2](410,63)(503,123)
    \GluonArc[clock](449.062,97.5)(20.439,-126.169,-347.281){3.25}{8}
    \Text(47,39)[lb]{\Large{\Black{$t^{(2n)}_L$}}}
    \Text(85,86)[lb]{\Large{\Black{$g^{(n)}$}}}
    \Text(127,56)[lb]{\Large{\Black{$t^{(n)}_R$}}}
    \Text(83,26)[lb]{\Large{\Black{$t^{(n)}_L$}}}
    \Text(146,12)[lb]{\Large{\Black{$H^{(0)}$}}}
    \Text(143,103)[lb]{\Large{\Black{$t^{(0)}_R$}}}
    \Text(185,41)[lb]{\Large{\Black{$t^{(2n)}_L$}}}
    \Text(228,109)[lb]{\Large{\Black{$g^{(n)}$}}}
    \Text(224,41)[lb]{\Large{\Black{$t^{(n)}_L$}}}
    \Text(264,42)[lb]{\Large{\Black{$t^{(0)}_L$}}}
    \Text(315,85)[lb]{\Large{\Black{$t^{(0)}_R$}}}
    \Text(312,41)[lb]{\Large{\Black{$H^{(0)}$}}}
    \Text(459,36)[lb]{\Large{\Black{$H^{(0)}$}}}
    \Text(427,60)[lb]{\Large{\Black{$t^{(2n)}_R$}}}
    \Text(464,79)[lb]{\Large{\Black{$t^{(n)}_R$}}}
    \Text(424,123)[lb]{\Large{\Black{$g^{(n)}$}}}
    \Text(493,102)[lb]{\Large{\Black{$t^{(0)}_R$}}}
    \Line[dash,dashsize=10,arrow,arrowpos=0.5,arrowlength=5,arrowwidth=2,arrowinset=0.2](121,28)(181,29)
    \Line[arrow,arrowpos=0.5,arrowlength=5,arrowwidth=2,arrowinset=0.2](123,97)(176,97)
    \Text(372,41)[lb]{\Large{\Black{$t^{(2n)}_L$}}}
  \end{picture}
}
\label{diagrams}
\end{center}
\caption{\sf The dominant diagrams in the unitary gauge generating an effective
$t^{(2n)}_L t^{(0)}_R H^{(0)}$ coupling. }
\end{figure}
%%%%%%%%%%%%%%%%%%%%%%%%%%%%%%%%%%%%%%%%%%%%%%%%%%%%%%%%%%%%%%%%%%% 
As seen from eq. (\ref{KKquarks}), at any KK-level $n$, excepting
$n = 0$, there are four top-quark excitations: $t^{(n)}_L,
T^{(n)}_R, T^{(n)}_L$ and $t^{(n)}_R$, the first two being
members of electroweak $SU(2)$ doublets while the last two are
singlets. For the zero-modes there is no right-handed doublet
member, $T^{(0)}_R$, nor a left-handed singlet, $T^{(0)}_L$.

The effective coupling which we wish to calculate involves a
decay of a $2n$-level top quark to a zero mode top quark and a zero
mode Higgs scalar. The $SU(2)$ doublet nature of the Higgs boson
and the nonexistence of $T^{(0)}_R$ and 
$T^{(0)}_L$ leaves only
the following possibilities $t^{(2n)}_R t^{(0)}_L H^{(0)}$ and
$t^{(2n)}_L t^{(0)}_R H^{(0)}$. 

The four-dimensional theory with the tower of Kaluza-Klein states
is valid up to the cut-off scale $\Lambda$. The
magnitude of a coupling at $\Lambda$ is determined by the theory
which takes over beyond this energy and is to be regarded as a
boundary condition for mUED. A common practice, pioneered, as
noted earlier, in the context of masses of KK-states in  minimal
UED \cite{mUED}, is to take this boundary value of the coupling at
$\Lambda$ to be zero and obtain its magnitude at low energy
through calculable corrections. We evaluate the KK-number
non-conserving couplings using the same principle.

In this section we present some details of the calculation which
is performed in the
unitary gauge\footnote{We have verified that identical results
are obtained in the `t Hooft-Feynman gauge.}.  The dominant
contributions to the first of these couplings\footnote{The
$t^{(2n)}_R t^{(0)}_L H^{(0)}$ coupling is obtained from similar
diagrams -- with $(L \leftrightarrow R)$ exchange -- which we
have not shown.} will arise from the Feynman diagrams shown in
Fig. 1. We ignore smaller contributions which are generated, for
example, by virtual $W^{\pm(1)}$ exchange.

Each of the diagrams Fig. 1(a), 1(b), and 1(c) are individually
divergent. We use dimensional regularisation to evaluate them.
Using the techniques of \cite{velt} the contributions can be
expressed after euclideanisation in terms of scalar loop
integrals which include the divergent pieces:

\begin{equation}
\frac{i}{\pi^2} \int d^nq \frac{1}{\left[q^2 + m^2\right]} = m^2(-\Delta - 1 +
\ln m^2) \;\;,
\label{d1} 
\end{equation}

\begin{equation}
\frac{i}{\pi^2} \int d^nq \frac{1}
{\left[q^2 + m^2\right]\left[(q + p)^2 + m^2\right]} = \Delta +
{\rm ~finite ~terms}\;\;,
\label{d2} 
\end{equation}
where
\begin{equation}
\Delta = -\frac{2}{n-4} + \gamma - \ln \pi \;\;,\;\; \gamma =
~{\rm Euler's ~constant} \;\;. 
\end{equation}
In Pauli-Villars  regularisation, the momentum integral in eq.
(\ref{d1}) is quadratically divergent while the one in eq.
(\ref{d2}) has a logarithmic behaviour.

In presenting the contributions from the diagrams in Fig. 1 we
encapsulate the couplings in  a common factor:
\begin{equation}
\xi = -\left(\frac{g_3^2}{16 \pi^2}\right) \frac{m_t}{v}
\left(T^c_{ab} T^c_{ba} \right) .
\label{xi}
\end{equation}

Using eqs. (\ref{d1}) and (\ref{d2}) we find for the
contribution from Fig. 1(a) to be
\begin{eqnarray}
-i {\cal M}_1 &=& \xi ~\bar{u}_0(k) \left\{ -\frac{1}{M_n^2}
\left[M_n^2(-\Delta - 1 +
\ln M_n^2)\right]  \right. \nonumber \\
&+& \left. \Delta \left(4 - \frac{1}{M_n^2}\left[ 2 M_n^2
-\frac{3}{2} M_{2n}^2\right] \right)  + {\rm ~finite } \right\}
\frac{1 - \gamma_5}{2} u_2(p) \;\;.
\label{c1}
\end{eqnarray}

Above,  $p$ and $k$ are the four momenta of the $t^{(2n)}_L$ and 
$t^{(0)}_R$.  Similarly from Figs. 1(b) and 1(c) we respectively get
\begin{eqnarray}
-i {\cal M}_2 &=& \xi ~\bar{u}_0(k) \left\{ \frac{M_{2n}^2}{M_{2n}^2 -
M_0^2}\frac{1}{M_n^2} \left[M_n^2(-\Delta - 1 + \ln M_n^2)\right]
\right. \nonumber \\
&+& \left.  \Delta \frac{1}{M_{2n}^2 - M_0^2}\left(-3 M_{2n}^2 +
\frac{1}{M_n^2}\left[\frac{1}{2} M_n^2 M_{2n}^2 -\frac{3}{2} M_{2n}^4\right] + {\rm
~finite} \right)  \right\}  \frac{1 -
\gamma_5}{2} u_2(p) \;\;,
\label{c2}
\end{eqnarray}
and
\begin{eqnarray}
-i {\cal M}_3 &=& \xi ~\bar{u}_0(k) \left\{ \frac{M_0^2}{M_0^2 -
M_{2n}^2}\frac{1}{M_n^2} \left[M_n^2(-\Delta - 1 + \ln M_n^2)\right]
\right. \nonumber \\
&+& \left.  \Delta \frac{1}{M_0^2 - M_{2n}^2}\left(-3 M_0^2 +
\frac{1}{M_n^2}\left[-\frac{1}{2} M_n^2 M_0^2 -\frac{3}{2} M_0^4\right] + {\rm
~finite} \right)  \right\}  \frac{1 -
\gamma_5}{2} u_2(p) \;\;.
\label{c3}
\end{eqnarray}

The leading (quadratic)
divergences cancel out when eqs.  (\ref{c1}) - (\ref{c3}) are
taken together. As remarked earlier, in  the spirit of mUED
calculations  the boundary value of the effective
$t_L^{(2n)}t_R^{(0)}H^{(0)}$ coupling is taken as zero at the
scale $\Lambda$.  At lower energies, $\mu$, the net contribution
is  logarithmically dependent on the energy scale -- i.e.,
proportional to $\ln ({\Lambda/\mu})$ and one gets from eqs.
(\ref{c1}) - (\ref{c3}):
\begin{eqnarray}
g^{\rm eff}_{t_L^{(2n)}t_R^{(0)}H^{(0)}} &=& \xi  \ln
\left(\frac{\Lambda}{\mu}\right)  \left\{1 +
\frac{1}{M_n^2}\left[M_n^2 \left(-2 + \frac{1}{2}\frac{M_{2n}^2 +
M_0^2}{M_{2n}^2 - M_0^2} \right) + \frac{3}{2} \left(M_{2n}^2 - (M_{2n}^2
+ M_0^2) \right) \right] \right\}   \frac{1 - \gamma_5}{2} 
\nonumber \\
&=& -\frac{1}{2} ~\xi  \ln \left(\frac{\Lambda}{\mu}\right)
\frac{1 - \gamma_5}{2} \;\;,
\label{coup}
\end{eqnarray}
where in the last step we have substituted $M_n = n/R$ for all
$n$. Notice that the resultant coupling is independent of $n$.

\section{Decays of a $2n$-level top quark}

We now turn to an examination of the decay rate of a $2n$-level KK top
quark state induced through the coupling calculated in the previous
section. We also compare it with other KK-number conserving
decays that are allowed but are phase space suppressed.

In general for a heavy fermion $F$ of mass $m_F$ decaying to a
different fermion $f$ and a scalar $h$ with masses $m_f$ and
$m_h$ respectively the decay width is
\begin{equation}
\Gamma(F \rightarrow f h) = \frac{\tilde{g}^2}{8 \pi m_F^3} \left[
(m_F - m_f)^2 - m_h^2 \right] \left\{ \left(m_F^2 - m_f^2 -
m_h^2\right)^2 - 4 m_h^2 m_f^2 \right\}^{1/2} \;\;.
\label{width}
\end{equation}
Above, $\tilde{g}$ is the strength of the effective Yukawa coupling
between $F, f$, and $h$.

For the case at hand using eq. (\ref{coup}) we then have
\begin{equation}
\Gamma\left(t_L^{(2n)} \rightarrow t_R^{(0)}H^{(0)}\right) =
\left[\frac{1}{2} \xi  \ln \left(\frac{\Lambda}{\mu}\right)
\right]^2 \left(\frac{2n/R}{8 \pi} \right) \;\;, 
\label{width00}
\end{equation}
where we have ignored the zero-mode masses compared to $2n/R$.
The mass scale $\mu$ has to be identified here with $m_F = 2n/R$.

This decay rate is to be compared with the KK-number conserving
decays which proceed via tree-level couplings.  As a typical
example we can consider the decay $t_L^{(2n)} \rightarrow
t_R^{(n)}H^{(n)}$. Here the coupling strength is simply $m_t/v$.
This decay would have been forbidden by phase space considerations
but for the mUED corrections to the KK-state masses. Keeping only
the strong interaction effects for illustration\footnote{For the
numerical results in the following section we keep full mUED
corrections \cite{mUED}.}  the corrected mass
$\bar{m}_n$ of the $n$-th KK quark state is given by \cite{mUED}
\begin{equation}
\bar{m}_n = m_n \left[ 1 + 3 ~\frac{g_3^2}{8 \pi^2}
~\ln\left(\frac{\Lambda}{\mu}\right)  \right] \;\;.
\end{equation}
This correction has the same form for quarks of both chirality.
Obviously, the Higgs scalar and its excitations receive no
corrections from the strong interactions. Substituting the above
in eq. (\ref{width}) one has 
\begin{equation}
\Gamma\left(t_L^{(2n)} \rightarrow t_R^{(n)}H^{(n)}\right) =
\left[\frac{m_t}{v} \right]^2  ~\ln \left(\frac{\Lambda}{\mu}\right)
 ~\left(\frac{n/R}{16 \pi} \right) \;\;. 
\label{widthnn}
\end{equation}
The decay width for more general possibilities such as 
$t_L^{(2n)} \rightarrow t_R^{(m)}H^{(2n-m)}$ can be readily
obtained using the appropriate product particle masses in eq. (\ref{width}).

%%%%%%%%%%%%%%%%%%%%%%%%%%%%%%%%%%%%%%%%%%%%%%%%%%%%%%%%%%%%%%%%%%% 
\begin{figure}[thb] 
\begin{center} 
\includegraphics[scale=0.3,angle=0]{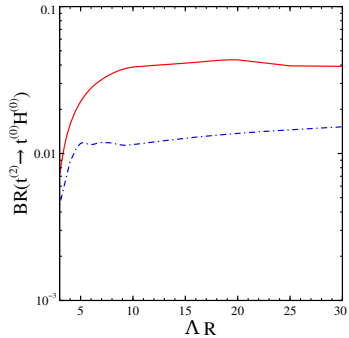}
\caption{\sf The branching ratio for the $t^{(2)}
\rightarrow t^{(0)}H^{(0)}$ mode as a function of $\Lambda R$ using
the full calculation of $t^{(2)}$ decay. The red
solid (blue dot-dashed) curve is for $t_L^{(2)}$ ($t_R^{(2)}$)
decay.}
\label{BR} 
\end{center} 
\end{figure} 
%----------------------

From eqs. (\ref{xi}), (\ref{width00}), and (\ref{widthnn}) we obtain
\begin{equation}
\frac{\Gamma\left(t_L^{(2n)} \rightarrow t_R^{(0)}H^{(0)}\right)}
{\Gamma\left(t_L^{(2n)} \rightarrow t_R^{(n)}H^{(n)}\right)} =
\left[ \left(\frac{g_3^2}{16 \pi^2}\right) 
\left(T^c_{ab} T^c_{ba} \right) \right]^2
  \ln \left(\frac{\Lambda}{\mu}\right) = 
\left[ 3\left(\frac{\alpha_s}{4 \pi}\right)  \right]^2
  \ln \left(\frac{\Lambda R}{2n}\right)\;\;. 
\label{br}
\end{equation}

The current practice is to choose $\Lambda$ such that
$\Lambda R \sim 10$. Masses of KK-states must not exceed
$\Lambda$ which implies that the above formulation is
meaningful for $n \leq 5$.
It bears mention that  the branching ratio in eq.
(\ref{br}) tends to zero as $2n \rightarrow \Lambda R$.

Our interest in the next section will be to examine the
possibility of detection of the KK-number non-conserving decay of
second level KK top quarks after their pair production at the LHC. 
This decay has to compete with the KK-number conserving decays.
We find that the dominant decay modes of the latter type are $t_L^{(2)}
\rightarrow W^{+(1)} b_L^{(1)}, W^{+(2)} b_L^{(0)}, W^{3(1)} t_L^{(1)},
h^{0(1)} t_R^{(1)}, B^{(1)} t_L^{(1)}$ and $t_R^{(2)}
\rightarrow h^{+(2)} b_R^{(0)}, h^{0(1)} t_L^{(1)}, B^{(1)} t_R^{(1)},
 h^{+(1)} b_L^{(1)}$. The branching ratio for the decay  $t^{(2)}
\rightarrow t^{(0)}H^{(0)}$ taking into account all the KK-number
conserving decay modes is shown in Fig. \ref{BR} as a function of
the parameter $\Lambda R$. The red solid curve corresponds to the
decay of a $t_L^{(2)}$ quark while the blue dot-dashed curve is
for $t_R^{(2)}$ decay.

\section{Detection prospect of the second-level KK top-quark}

In this section we discuss how the
$t_L^{(2)}t_R^{(0)}H^{(0)}$ coupling can be experimentally
probed with particular reference to the LHC. We consider the pair
production of  $t_{L,R}^{(2)} \bar{t}_{R,L}^{(2)}$ at the LHC and
the subsequent decay of both of them through the
$t_{L,R}^{(2)}t_{R,L}^{(0)}H^{(0)}$ coupling and compare this
signal with the Standard Model (SM) background\footnote{Below we
consider the signal due to the production of a $t_{L}^{(2)}$
along with a $\bar{t}_{R}^{(2)}$. Inclusion of $t_{R}^{(2)}
\bar{t}_{L}^{(2)}$ production will enhance the signal by a factor
of 2.}. Assuming that
both second-level top-quarks decay in the $t^{(0)}H^{(0)}$ mode the
signal consists of two top quarks\footnote{Obviously, one would
be a top anti-quark but we forego this distinction for ease of
presentation.}  and two Higgs bosons such that the correct
pairing leads to identical invariant masses for the two
$t^{(0)}H^{(0)}$ pairs.   We estimate the Standard Model
background for this channel and find it to be insignificant.
However, with $\sqrt{s} = 13$ TeV  and an integrated
luminosity of 300 fb$^{-1}$ the signal is small in number and
inadequate for vindicating the strength of the coupling. 
On the other hand, with the HL-LHC option at the same $\sqrt{s}$ with $\int
{\cal L} dt$ = 3000 fb$^{-1}$ the signal could be viable.  For the
HE-LHC with $\sqrt{s} = 33$ TeV and $\int {\cal L} dt$ = 300
fb$^{-1}$ the reach would be more. The 100 TeV hadron FCC would
obviously do the best. 

%%%%%%%%%%%%%%%%%%%%%%%%%%%%%%%%%%%%%%%%%%%%%%%%%%%%%%%%%%%%%%%%%%% 
\begin{figure}[thb] 
\begin{center}
\includegraphics[scale=0.3,angle=0]{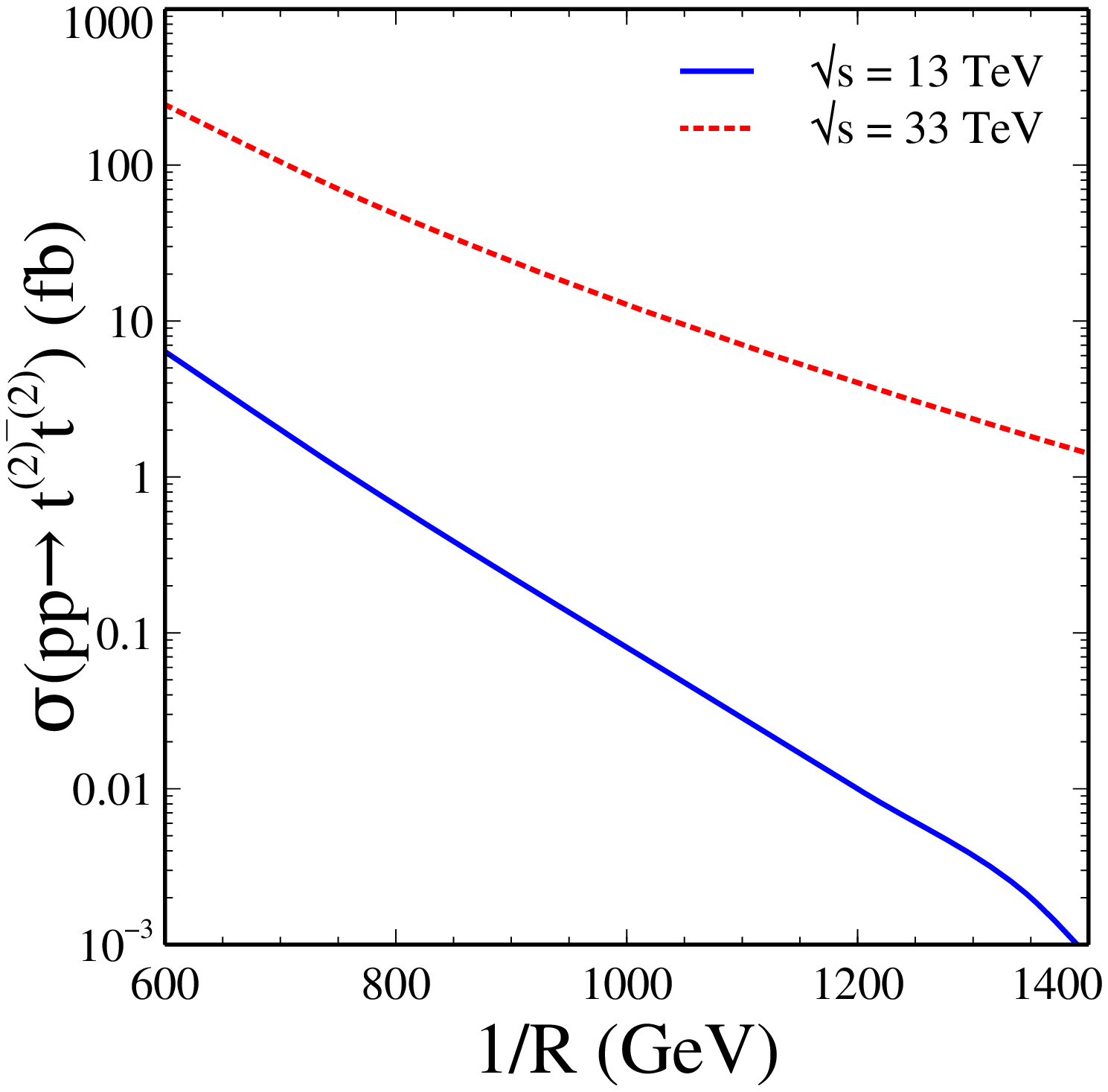}
\caption{\sf The production cross section for a $t^{(2)}
\bar{t}^{(2)}$ pair at the LHC. The blue solid (red dashed) curve
corresponds to $\sqrt{s}$ = 13 TeV (33 TeV).} 
\label{LHCprod} 
\end{center} 
\end{figure} 
%----------------------

We use the CalcHEP implementation of mUED \cite{mUEDcalc,
calchep} to generate the events. A parton-level Monte Carlo has
been utilized with the CTEQ6l \cite{cteq} distribution functions.
The renormalization scale (for $\alpha_s$) and the factorisation
scale (for the parton distributions) are both taken\footnote{We
have checked that if this scale is chosen as $4/R$ -- to account
for the production of two  $t^{(2)}$ states, each of mass $2/R$
--  the production cross section is enhanced by about 20\%.} as
$2/R$.

The production of the  $t^{(2)}
\bar{t}^{(2)}$ pair proceeds through gluon-gluon fusion -- both
$s$-channel and $t$-channel processes -- as well as $q \bar{q}$
annihilation. We find that at the $\sqrt{s}$  that we study the
former dominate. The production cross sections for LHC running at
$\sqrt{s}$ = 13 TeV and in the future at a 33 TeV  HE-LHC
are shown in Fig. \ref{LHCprod}. 

%---------------------------------------------------------------------------
\begin{figure}[tb] 
\begin{center} 
{\includegraphics[scale=0.4,angle=0]{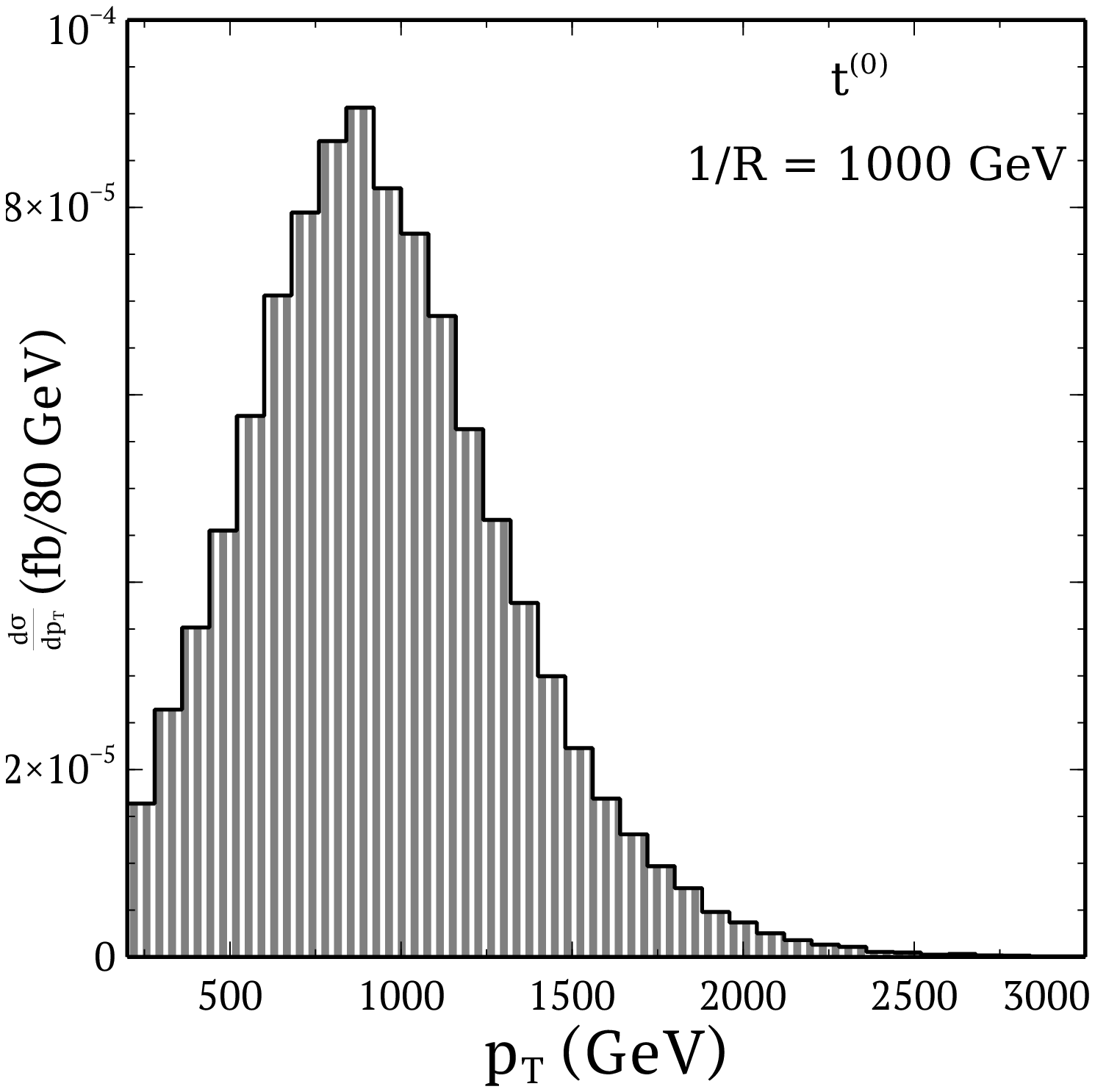}
\includegraphics[scale=0.4,angle=0]{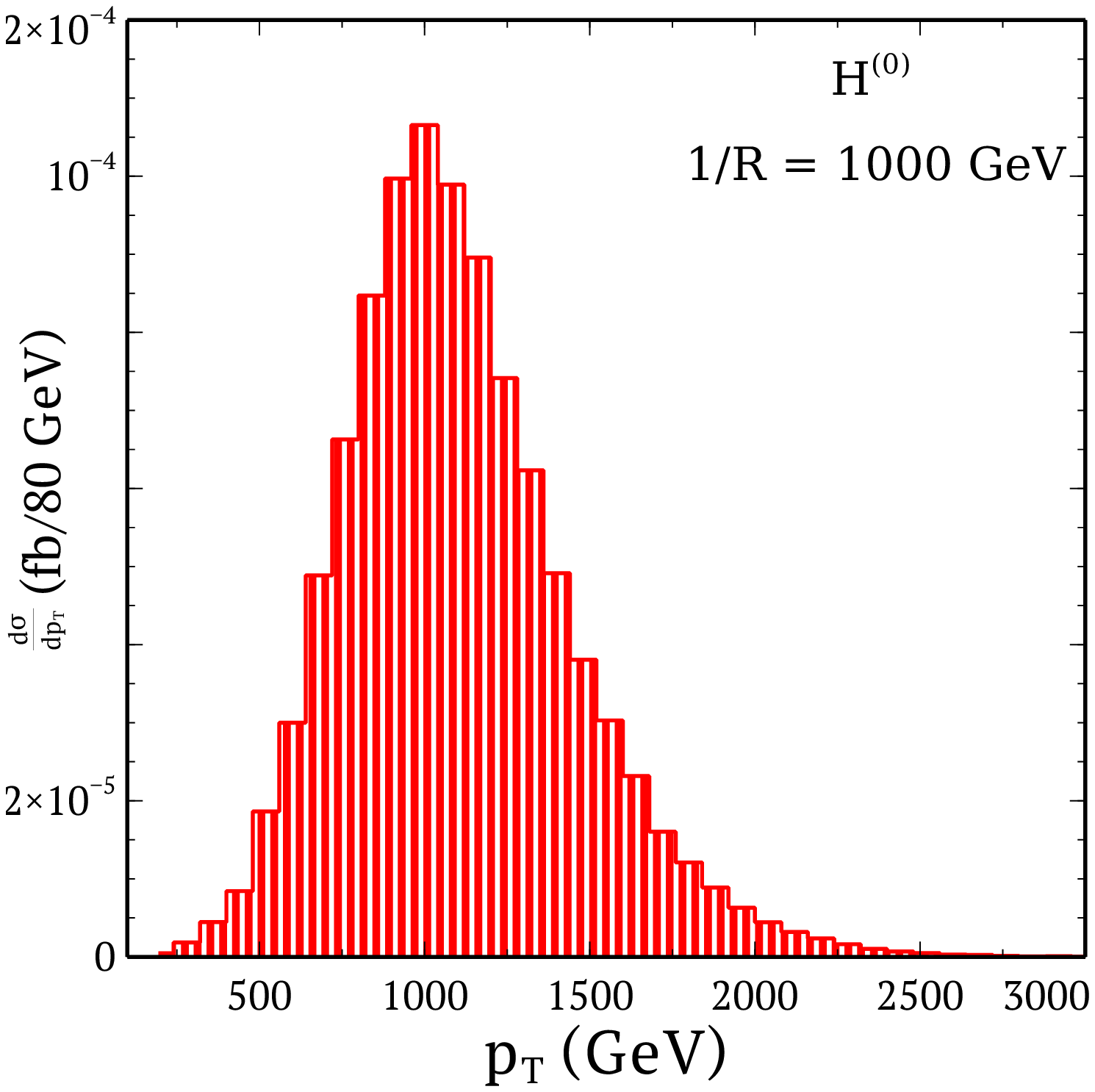}}
\caption{\sf \small The $p_T$ distributions of the top quark
(left) and the Higgs boson (right)  for the signal at the LHC
running at $\sqrt{s}$ = 13 TeV.  }
\label{ptdist} 
\end{center} 
\end{figure} 
%---------------------------------------------------------------------------

The goal of this section is only to make a preliminary
examination of this channel.  So, we have refrained from
including detailed detector simulation or indeed the subsequent
decays of the top-quark or the Higgs bosons. We incorporate these
effects by appropriate detection efficiency factors for these
particles after applying kinematic cuts discussed later.  

Since the $t^{(2)}$ states have a mass more than at least 1
TeV, the top quark and Higgs boson produced in their decays are
highly boosted.  Their further decay products are boosted in the
direction of motion of the parent particle which results in `fat'
jets for the top quark and the Higgs boson which have a
substructure consisting of subjets of b-quarks and light quarks
and/or leptons. Since in this work we are not delving into this
detailed substructure we regard the signal event as consisting of
two $t^{(0)}$ and two $H^{(0)}$ fat jets. A characteristic
measure of the `fatness' is the opening angle parameter
\begin{equation}
\Delta R = \sqrt{(\Delta \eta)^2 + (\Delta \phi)^2} \;\;,
\end{equation}
where $\eta$ is the pseudorapidity and $\phi$ the azimuthal
angle. A common practice is to take  $\Delta R \sim 2 m/p_T$ where $m$
is the mass of the particle \cite{fatjet}. As a typical example,
we show in Fig. \ref{ptdist} the $p_T$ distributions for the top
quark and one of the Higgs bosons in the signal for $1/R =$ 1000
GeV and $\sqrt{s}$ = 13 TeV. It is seen that both distributions
peak near 1 TeV and are small for $p_T <$ 500 GeV.  Therefore the
four fat jets from the signal events can be expected to have
$\Delta R$ around 0.35 for the top jets and 0.25 for the Higgs jets.
For $1/R$ = 800 GeV the $p_T$ distribution of the fat jets is
peaked at a slightly lower value ($\sim$ 800 GeV). We have
verified that at $\sqrt{s}$ = 33 TeV these results are hardly
affected. So in all cases of interest the `fat' jets have $\Delta
R \sim 0.30$.

For the signal as
well as the background for  the four jets we impose the
following $p_T$ and pseudorapidity  cuts:
\begin{equation}
p_T > 25 {\rm ~GeV}\;\;,  \;\; |\eta| < 2.5 \;\;.
\end{equation}
In addition, all four  fat jets are required to be isolated.
In view of our previous discussion, for any two
of them $i,j$ we require:
\begin{equation}
\Delta R_{ij} = \sqrt{(\Delta \eta)_{ij}^2 + (\Delta
\phi)_{ij}^2} > 0.5 \;\;.
\end{equation}
From the surviving events we pick those for which there are
two distinct $t^{(0)}H^{(0)}$ pairs of the same invariant mass.
We ensure that the $p_T$ of the  two reconstructed $t^{(2)}$
are balanced  to within 10\%.  %
%---------------------------------------------------------------------------
\begin{figure}[tb] 
\begin{center} 
{\includegraphics[scale=0.4,angle=0]{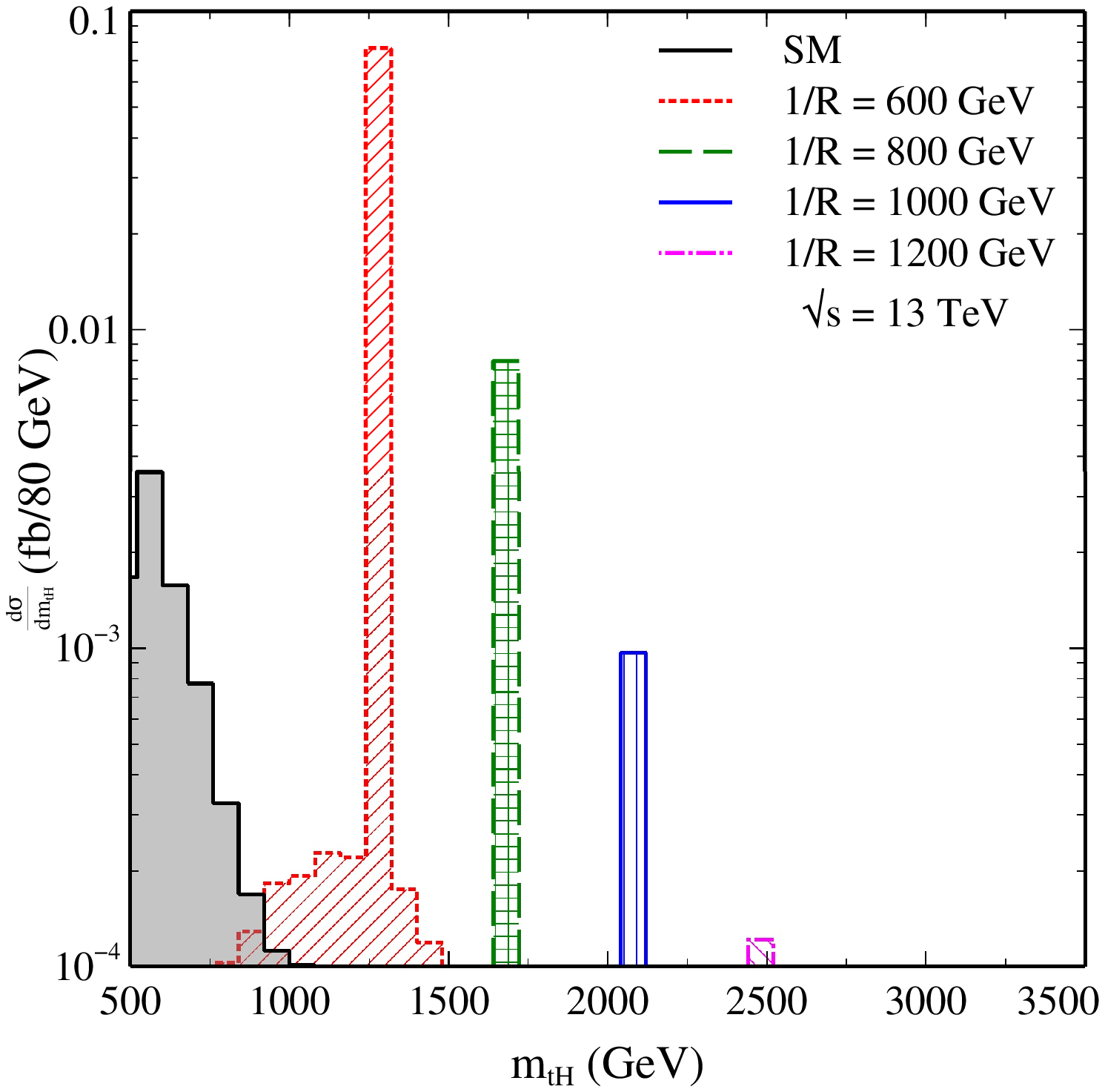}
\includegraphics[scale=0.4,angle=0]{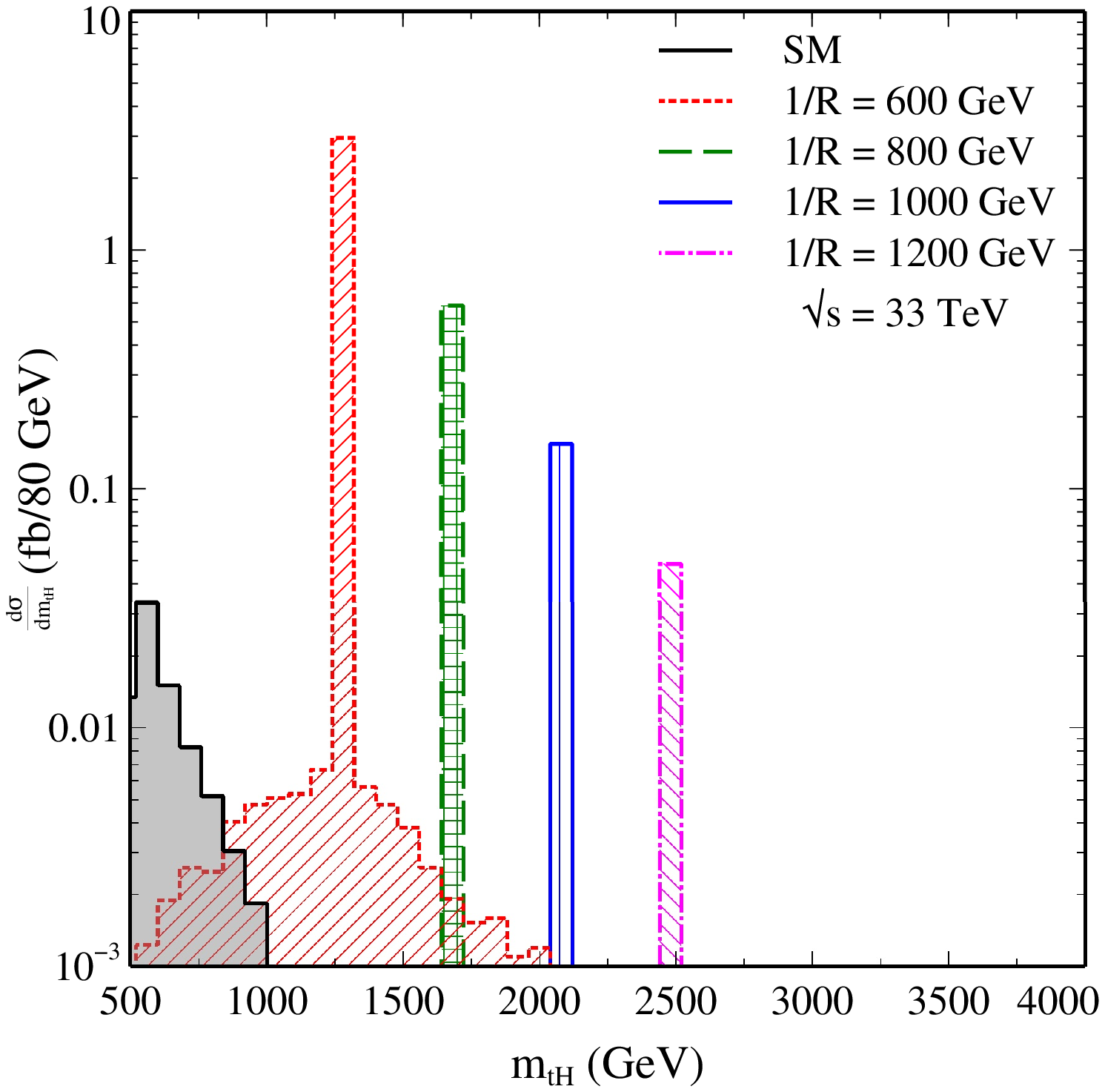}}
\caption{\sf \small The cross section for the $(tH)(tH)$ signal
at the LHC as a function of the $t^{(0)}H^{(0)}$ invariant mass.
The histograms are for the signal at the LHC running at $\sqrt{s}$ = 13
TeV (left) and 33 TeV (right) for different choices of $1/R$
(explained in the legend). The SM background is shown shaded gray
in both panels. }
\label{tHtH} 
\end{center} 
\end{figure} 
%---------------------------------------------------------------------------

In Fig. \ref{tHtH} is shown the cross section for the above
process as a function of the $t^{(0)}H^{(0)}$ invariant mass. In
the left (right) panel are the results for $\sqrt{s}$ = 13 (33)
TeV. The histograms correspond to the signal for $1/R$ = 600 GeV
(red dotted), 800 GeV (green dashed), 1000 GeV (blue
solid), and 1200 GeV (pink dot-dashed).  For both
panels the SM background, shown shaded gray, is insignificant in the
region of the signal. So, a signal of 10 events would be  strong
evidence for this model. 

The detection efficiency of boosted top quarks and Higgs bosons
have been under much investigation in the literature.  Using jet
substructure features the tagging efficiency of boosted top
quarks with $p_T$ in the 800 - 1000 GeV range  decaying
hadronically, i.e., with a branching ratio 2/3, is estimated
around $\epsilon_{top}$ = 0.40-0.45 \cite{topeff}. For a boosted
Higgs boson similar analyses yield an efficiency of
$\epsilon_{h\rightarrow b\bar{b}}$ = 0.94 for the $b\bar{b}$
decay mode \cite{higgseff} which has a branching ratio of about
60\%.

As seen from the left panel of Fig. \ref{tHtH}, for
$\sqrt{s}$ = 13 TeV with 300 fb$^{-1}$ integrated
luminosity\footnote{The expected  energies and luminosities
of future $pp$-colliders used here are from
\cite{futurepp}.}  the
detection is unlikely.  For the lowest $1/R$ that we consider,
namely 600 GeV, one has around 30 events.  Using the
above-mentioned top quark and Higgs boson tagging
efficiencies\footnote{We conservatively include only the
$b\bar{b}$ decay mode of the Higgs.} one is left with the signal
of $((2/3) \epsilon_{top})^2 (0.6 \epsilon_{h\rightarrow
b\bar{b}})^2 \times 30 \sim$ 1 event only. For the high
luminosity  HL-LHC option ($\int {\cal L} dt$ = 3000
fb$^{-1}$) this will become a healthy 10-event signal. However,
with $1/R$ = 800 GeV the signal will fall to around 1 event.  On
the other hand, at a  HE-LHC with $\sqrt{s}$ = 33 TeV (right
panel of Fig. \ref{tHtH}) the signal is enhanced roughly by two
orders of magnitude and could remain viable till $1/R$ = 1 TeV
with $\int {\cal L} dt$ = 3000 fb$^{-1}$.  We have checked that
with a 100 TeV hadron FCC even for 100 fb$^{-1}$ integrated
luminosity this reach would go up to $1/R$ = 2.5 TeV
for which we find 10 events.

\section{Summary and Conclusions}

In this work we have calculated the coupling  of a 2$n$-level KK
top-quark to a zero-mode top and a zero-mode Higgs boson in the universal
extra-dimensional model. Such a coupling violates KK-number but
respects KK-parity and is induced by loop diagrams. The dominant
contribution comes from $n$-level quark and gluon mediation. We
evaluate this coupling and show that it is independent of $n$.

We use this coupling to estimate the branching ratio of a second
level KK-top quark for this KK-number non-conserving mode, which
has the advantage of a large phase space. Considering the pair
production of such second level top quarks at the LHC with
$\sqrt{s}$ = 13 TeV  and 33 TeV (HE-LHC) we examine the prospects of the
detection of both of them in this decay mode. Our results are
encouraging for the High Luminosity or High Energy runs of the
LHC. A hadron FCC with $\sqrt{s}$ = 100 TeV would considerably
expand the reach of this program.

{\bf Acknowledgements:} U.K.D. is grateful to Shankha Banerjee
and Tanumoy Mandal for many helpful discussions. U.K.D. is
supported by  funding from the Department of Atomic Energy,
Government of India for the Regional Centre for Accelerator-based
Particle Physics, Harish-Chandra Research Institute (HRI).  A.R.
is partially funded by  the Department of Science and Technology
Grant No. SR/S2/JCB-14/2009.
 
{\Large{\bf Appendix: Feynman rules}}
\setcounter{equation}{0}  
\setcounter{section}{1}  

\renewcommand{\thesection}{\Alph{section}}
\renewcommand{\theequation}{\thesection-\arabic{equation}}  

Here we list the Feynman rules relevant for our calculation.
$i,j$ are colour indices. In the first two vertices
the chirality index is suppressed while in the third the colour
index is not shown.

\fcolorbox{white}{white}
{  \begin{picture}(244,166) (15,-91)
    \SetWidth{1.0}
    \SetColor{Black}
    \Photon(20,-10)(70,-10){4}{8}
    \Line[arrow,arrowpos=0.5,arrowlength=5,arrowwidth=2,arrowinset=0.2](70,-10)(120,30)
    \Line[arrow,arrowpos=0.5,arrowlength=5,arrowwidth=2,arrowinset=0.2](120,-50)(70,-10)
    \Text(40,-80)[lb]{\Black{$\equiv -i g_{s} \gamma^{\mu} (\lambda_a)_{ij} $}}
    \Text(39,0)[lb]{\Black{$g^{(n)}_a$}}
    \Text(110,-45)[lb]{\Black{${t}^{(2n)}_{j}, {T}^{(2n)}_{j}$}}
    \Text(110,10)[lb]{\Black{$t^{(n)}_{i},{T}^{(n)}_{i}$}}

    \Photon(180,-10)(230,-10){4}{8}
    \Line[arrow,arrowpos=0.5,arrowlength=5,arrowwidth=2,arrowinset=0.2](230,-10)(280,30)
    \Line[arrow,arrowpos=0.5,arrowlength=5,arrowwidth=2,arrowinset=0.2](280,-50)(230,-10)
    \Text(200,-80)[lb]{\Black{$\equiv -i g_{s} \gamma^{\mu} (\lambda_a)_{ij}$}}
    \Text(199,0)[lb]{\Black{$g^{(n)}_a$}}
    \Text(270,-45)[lb]{\Black{${t}^{(n)}_{j}, {T}^{(n)}_{j}$}}
    \Text(270,10)[lb]{\Black{$t^{(0)}_{i},{T}^{(0)}_{i}$}}

    \Line[dash,dashsize=5,arrow,arrowpos=0.5,arrowlength=5,arrowwidth=2,arrowinset=0.2](390,-10)(340,-10)
    \Line[arrow,arrowpos=0.5,arrowlength=5,arrowwidth=2,arrowinset=0.2](390,-10)(440,30)
    \Line[arrow,arrowpos=0.5,arrowlength=5,arrowwidth=2,arrowinset=0.2](440,-50)(390,-10)
    \Text(360,-80)[lb]{\Black{$\equiv -i m_t/v$}}
    \Text(359,0)[lb]{\Black{$H^{(0)}$}}
    \Text(430,-45)[lb]{\Black{${t}^{(n)}_{L,R}, {T}^{(n)}_{L,R}$}}
    \Text(430,8)[lb]{\Black{$t^{(n)}_{R,L},{T}^{(n)}_{R,L}$}}

  \end{picture}
}

\end{document}